\newif\ifisonecolumn
\isonecolumntrue 

\ifisonecolumn
\documentclass[onecolumn,draftclsnofoot,12pt]{IEEEtran}
\else
\documentclass{IEEEtran}
\fi

\input{defi_let_ULDL.def}

\begin{document}

\TwoOneColumnAlternate
{
}
{\renewcommand{\baselinestretch}{1.37}
}

\title{A lower bound on the data rate of dirty paper coding in general noise and interference}
\author{Itsik~Bergel,~\IEEEmembership{Senior Member,~IEEE,}
        Daniel~Yellin,~\IEEEmembership{Senior Member,~IEEE,}
        and~Shlomo~Shamai,~\IEEEmembership{Fellow,~IEEE}
\thanks{\copyright {} 2015 IEEE. Personal use of this material is permitted. Permission from IEEE must be obtained for all other uses, in any current or future media, including reprinting/republishing this material for advertising or promotional purposes, creating new collective works, for resale or redistribution to servers or lists, or reuse of any copyrighted component of this work in other works.}
\thanks{I. Bergel is with Faculty of Engineering, Bar-Ilan University, 52900 Ramat-Gan, Israel;
   (e-mail: bergeli@biu.ac.il).}
\thanks{D. Yellin is with Marvell Semiconductor Israel, Azorim Park, Petach-Tikva, 49527, Israel; (e-mail: daniel.yellin@marvell.com).}
\thanks{S. Shamai (Shitz) is with the Department of Electrical Engineering, Technion-
Israel Institute of Technology, Technion City, Haifa 32000, Israel; (e-mail:
sshlomo@ee.technion.ac.il)}
\thanks{Published in the IEEE Wireless Communications Letters, DOI: 10.1109/LWC.2014.2323355.}}

\maketitle

\begin{abstract}
Dirty paper coding (DPC) allows a transmitter to send information to a receiver in the presence of interference that is known (non-causally) to the transmitter.
The original version of DPC was derived for the case where the noise and the interference are statistically independent Gaussian random sequences. More recent works extended this approach
to the case where the noise and the interference are mutually independent and at least one of them is Gaussian. In this letter we further extend the DPC scheme by relaxing the
Gaussian and statistical independence assumptions. We provide lower bounds on the achievable data rates in a DPC setting for the case of possibly dependent noise, interference and input signals. Also,
the interference and noise terms are allowed to have arbitrary probability distributions. The bounds are relatively simple, are phrased in terms of second-order statistics, and are
tight when the actual noise distribution is close to Gaussian.
\end{abstract}

\section{Introduction}
The exponential growth of wireless data traffic on one hand and the limited availability of appropriate spectrum on the other, generates a strong motivation to significantly improve the spectral
efficiency per unit area offered by wireless systems. One clear example is today's cellular networks, where the overall data traffic grew by more than an order of magnitude during just the past four years
\cite{ericsson2013ericsson,NorthAmerica2011LTE}, and this trend is expected to continue and intensify. The conventional approach of deploying more base stations per unit area is
no longer a valid cost-effective mechanism to supply the huge growth in traffic demand. Therefore, more advanced techniques such as cooperative transmission and interference mitigation at the transmitter side are being considered
(e.g., \cite{ramprashad2011cooperative}) and standardized (e.g., \cite{gpp36819}) for the next generation of cellular networks.

In 1983, 
Costa presented the dirty paper
coding scheme (DPC) \cite{costa1983writing}. This scheme allows {\em full} cancellation of interference at the transmitter side, and leads to interference free data rates. The DPC scheme attracted considerable interest in recent years,
and state-of-the-art DPC approaches are applicable for the case where the noise and the interference terms are mutually independent of each other and of the input signal and at least one of these
terms is Gaussian  \cite{cohen2002gaussian}--\nocite{zamir2002nested}\cite{erez2005capacity}. In this work we further relax the model assumptions and consider the application of DPC to the case of non-Gaussian channels where the noise and interference terms may have arbitrary probability distributions,
and may also be statistically dependent of each other and/or of the input signal. In addition to the obvious application
of DPC for channels with non-Gaussian noise and interference, other common applications include: interference cancellation with partial channel state information (CSI) (e.g. \cite{wang2006adaptive,bergel2012IEEEI}),
Uplink-Downlink capacity balancing via variable feedback rate \cite{bergel2013preparation}, water marking (e.g., \cite{cohen2002gaussian,chen2001quantization}),  and more.

We present a lower bound on the achievable rate in a DPC setting. The bound is based on choosing a Gaussian input distribution, and then proving that the effect of the interference can be completely eliminated, while the effect of the
noise is bounded by the effect of a Gaussian noise with the same second order statistics. In the case of uncorrelated noise, the bounding data rate is identical to the capacity of the AWGN channel with the same noise variance
(and without interference). This bound may also be viewed as an extension of the `Gaussian noise is the worst noise' principle (e.g., \cite{blachman1957communication}--\nocite{dobrushin1959optimum}\cite{mceliece1981information})
to the case of DPC. Yet, the arguments that prove the worst noise principle do not hold in the DPC case, and a different proof is provided. Furthermore, one needs to recall that even in the absence of interference, the Gaussian
noise is the worst noise only if the input distribution is Gaussian (as an example see \cite{shamai1992worst}). In the presence of interference the situation is even more intricate.

The reminder of this paper is organized as follows: Section \ref{sec: system model} presents the system model and briefly discuss the results of previous works. Section \ref{sec: Main results} presents our main results
and Section \ref{sec: proof of rate bound} details their proofs. Section \ref{sec: conclusions} contains our concluding remarks.

\section{System model \& Previous works}\label{sec: system model}
We consider the additive noise additive interference channel, given by:
\begin{IEEEeqnarray}{rCl}\label{d: channel model}
Y = \ch   \cdot X +Z +N.
\end{IEEEeqnarray}
where $X$ is the transmitted data symbol, $\ch $ is a random channel gain and $Z$ is an interference term with variance of $\sigma_Z^2$. It is assumed that the transmitter has non-casual knowledge of both the channel
gain, $\ch $, and the interference term, $Z$, while the receiver knows only the channel gain. We consider a general additive noise, $N$, which may depend on the transmitted symbol, $X$, and the interference, $Z$. The channel is assumed to be memoryless, and obeys the probability
law\footnote{In this work we consider the case of real valued signals. The results hold also for the complex
valued case with proper complex terms \cite{neeser1993proper}, but the factor $1/2$ in Equations (\ref{e: basic theorem}) and (\ref{e: uncorrelated bound}) should be removed.}:
\begin{IEEEeqnarray}{rCl}
f_{\ch,Z,N|X}(h,z,n|x)&=&f_{N|X,Z}(n|x,z)
 f_{Z}(z)f_{\ch}(h)
\end{IEEEeqnarray}
where the distribution of the interference, $f_{Z}(z)$, the distribution of the channel, $f_{\ch}(h)$, and the conditional distribution of the noise, $f_{N|X,Z}(n|x,z)$, are given, while the transmitter can choose the
conditional distribution of the signal given the known interference and the channel, to maximize the rate.

We assume throughout an independent and identically distributed (iid) setting, i.e., $\ch$, $Z$ and $N$ are assumed to be composed of sequences of iid random variables (given $X$). Note that the results can be trivially extended to
any power-limited ergodic distributions, by using an interleaving argument. 

\subsection{Previous works}
The achievable rates in the presence of known interference of general distribution is characterized by the classical Gel'fand-Pinsker formula \cite{gel1980coding}: \begin{IEEEeqnarray}{rCl}\label{e: costa rate defintion}
R=I(U;Y| \ch   )-I(U;Z| \ch   )
\end{IEEEeqnarray}
where $U$ is an auxiliary random variable. This rate is achievable using random coding and binning with iid signaling, where 
\begin{IEEEeqnarray}{rCl}
f_{U,X,\ch,Z,N}(u,x,h,z,n)&=&f_{N|X,Z}(n|x,z)f_{\ch}(h)f_{Z}(z)
\TCnl\TCee \TwoOneColumnAlternate{\cdot}{} f_{X|U,Z}(x|u,z,h)f_{U|Z}(u|z,h) . \IEEEeqnarraynumspace
\end{IEEEeqnarray}
But, the optimal setting of this auxiliary variable in general is yet unknown, and so is the corresponding maximal achievable rate.

The DPC scheme gave the optimal choice of the auxiliary variable for the case of statistically independent noise, $N$, and interference, $Z$, both Gaussian processes \cite{costa1983writing}. Later, it was extended to the case of Gaussian noise and arbitrary
interference \cite{zamir2002nested,erez2005capacity}. Additionally, for the case of Gaussian interference Cohen and Lapidoth \cite{cohen2002gaussian} showed that Gaussian noise gives a lower bound on the achievable rate for
all noise distributions that are statistically independent of the input and the interference.

Costa showed \cite{costa1983writing} that in the presence of additive Gaussian noise, the channel capacity is identical to the capacity of the channel without interference. Cohen and Lapidoth \cite{cohen2002generalized} termed
this property \textit{private-public equivalence} (PPE) and showed that it does not hold for any noise distribution. They even conjectured that the PPE property holds only for additive Gaussian noise.

To further emphasize the effect of non-Gaussian noise, Cohen and Zamir \cite{cohen2008entropy} gave an example in which the difference between the channel capacity with and without interference is not even bounded. They further
speculated that the DPC scheme is efficient only if the noise distribution is Gaussian. Yet, \cite{cohen2008entropy} addressed only channels with discrete inputs, and hence, the distribution of the input, $X$, could not be Gaussian. In this work we take a
different approach, and show that the capacity of the channel can be lower bounded when the input distribution is allowed to be Gaussian.

The presented lower bound is equal to the no-interference capacity only if the noise and the input symbols are jointly Gaussian. Hence, the presented bound does not contradict the PPE conjecture of Cohen and Lapidoth
\cite{cohen2002generalized}.

\section{Main results}\label{sec: Main results}
Defining the variance and correlation coefficient of general random variables, $V$ and $W$ as $
\sigma_{V}^2=E\left[\left(V -E[V ]\right)^2\right]$
and 
\begin{IEEEeqnarray}{rCl}
\rho_{V W}=\frac{E\left[(V-E[V]) \left(W-E[W]\right)\right]}{\sigma_{V}\cdot\sigma_{W}}
\end{IEEEeqnarray}
 respectively, a lower bound on the capacity of the additive noise additive interference channel is given by the following theorem.
\begin{theorem}\label{Lem: rate lower bound}
Let $X_\tG$ be a Gaussian random variable with zero mean and variance of $\sigma_X^2$, and denote its distribution by $f_{X_\tG}(x)$. Define also the random variable $N_\tG$,
so that $(X_\tG,N_\tG,Z)$ follow the joint distribution: $f_{X_\tG,N_\tG,Z}(x,n,z)=f_{N|X,Z}(n|x,z) f_{Z}(z) f_{X_\tG}(x)$.

The capacity of the channel presented in (\ref{d: channel model}), subject to the average power constraint: $E[X^2] \le \sigma_X^2$, is lower bounded by:
\begin{IEEEeqnarray}{rCl}\label{e: basic theorem}
R
&\ge&
E_{ \ch   }\left[\frac{1}{2}
\log_2\left(1+\frac{\bigl(  \ch+\rho_{X_\tG N_\tG}\frac{\sigma_{N_\tG}}{\sigma_X}  \bigr)^2 }{1-\rho_{X_\tG N_\tG}^2-\rho_{Z N_\tG}^2}\cdot\frac{\sigma_{X}^2}{\sigma_{N_\tG}^2}\right)
\right].
\end{IEEEeqnarray}
\end{theorem}
\begin{IEEEproof} The proof is given in Section \ref{sec: proof of rate bound}.
\end{IEEEproof}

The main importance of Theorem \ref{Lem: rate lower bound} is its ability to lower bound the capacity of an arbitrary additive channel, based only on second order statistics of the relevant signals. Furthermore, the bound does not
depend on the interference variance and holds also for interference with unbounded variance. The theorem shows that correlation between the (known) interference and the noise is always beneficial and increases
the lower bound\footnote{In the presence of correlation between the (known) interference and the (unknown) noise, the noise can be decomposed into two parts. The first part is known (given the interference) while the second part
is unknown. The DPC scheme can mitigate the effect of the known part of the noise, and hence reduce the effect of the noise and increase the capacity}. On the other hand, correlation between the noise and the desired signal
can be beneficial or harmful, depending on the signs of the cross-correlation coefficient and the channel gain.

Note that the bound can be further tightened by adjusting the instantaneous transmission power according to the channel state.   The optimization of  the power allocation  scheme is a variation  of the well known water-pouring
solution \cite{cover2006elements}, which is somewhat more complicated due to the possible dependency of the noise power and the correlation coefficients on the instantaneous transmission power. The extension for variable power
allocation is straight-forward, and is therefore omitted.

 A simpler version of the bound is presented in Corollary \ref{cor: uncorrelated}, where we apply Theorem \ref{Lem: rate lower bound} to the case that the $X$ and $N$ are not correlated. 
\begin{corollary}[Uncorrelated noise]\label{cor: uncorrelated}
If, in addition to the conditions of Theorem \ref{Lem: rate lower bound}, the conditional distribution of the noise is such that $E\left[N|X=x\right]=E[N]$ for any $x$, then the capacity of the channel is lower bounded by:
\begin{IEEEeqnarray}{rCl}\label{e: uncorrelated bound}
R
&\ge&
E_{ \ch   }\left[\frac{1}{2}
\log_2\left(1+\frac{  \ch  ^2 \cdot \sigma_X^2 }{\sigma_{N_\tG}^2}\right)
\right].
\end{IEEEeqnarray}
\end{corollary}
\begin{IEEEproof} The corollary's condition leads to $E\left[X \left(N-E[N]\right)\right]=0$ for any distribution of $X$, and in particular $\rho_{X_\tG N_\tG}=0$. The proof follows immediately by substituting this zero cross-correlation into (\ref{e: basic theorem}) and noting that $\rho_{Z N_\tG}^2>0$ can only increase the bound.
\end{IEEEproof} 

Note that (\ref{e: uncorrelated bound}) is also a lower bound on the capacity of the interference free channel. The bound is tight if the noise is Gaussian, and the interference is statistically independent of the noise.

The condition in Corollary \ref{cor: uncorrelated} includes also the simpler case in which the noise is statistically independent of the transmitted signal. In such case, the noise variance is independent of the input signal, and the $\sigma_{N_\tG}^2$ term in (\ref{e: uncorrelated bound}) can be replaced by $\sigma_{N}^2$. If, in addition, the noise, $N$, and interference, $Z$, are
statistically independent, and at least one of them is Gaussian, then the bound in (\ref{e: uncorrelated bound}) converge to the results of \cite{cohen2002gaussian,zamir2002nested,erez2005capacity}.

Corollary \ref{cor: uncorrelated} is particularly useful when (\ref{e: uncorrelated bound}) is close to the interference free channel capacity. In such cases, the corollary shows that DPC is useful even in imperfect channel settings, and gives theoretical basis for the analysis of DPC in such cases (e.g., in broadcast channels with imperfect CSI).

The results above were derived using the following lemma, which is more general and allows the derivation of tighter bounds than (\ref{e: basic theorem}).

\begin{lemma}\label{Lem: general lower bound} The achievable rate in the channel of (\ref{d: channel model}) is lower bounded by
\begin{IEEEeqnarray}{rCl}\label{e: lemma equation}
R&\ge&\sup_{f_{X|\ch,Z}} E\Big[h(X| \ch=\eta,Z   )
\TwoOneColumnAlternate{ \\ \nonumber
&&}{}-\min_{\alpha,\beta}h\left( (1-\beta\eta)  X +(\alpha-\beta)Z -\beta N\big| \ch =\eta\right)  \Big]
\end{IEEEeqnarray}
where $h(\bullet)$ denotes the differential entropy and the distribution of $N$ is evaluated according to the chosen distribution of $X$. If the problem has constraints on the input distribution, then any tested distribution of $X$ must satisfy these constraints.
\end{lemma}
\begin{IEEEproof} See Section \ref{sec: proof of rate bound}.
\end{IEEEproof}

The lemma provides an easy way to derive expressions that lower bound the channel capacity. In particular, we note that any distribution of $X$ and any choice of the parameters $\alpha$ and $\beta $ that satisfy the
problem constraints will lead to a valid lower bound. A particularly interesting choice of parameters is given by $\alpha=\beta $, where the interference term in the right hand entropy  in (\ref{e: lemma equation}) is zeroed.
If in addition, the distribution of the input symbol is chosen so that $X$ is statistically independent of the known interference given $\ch$ ($f_{X|\ch,Z}=f_{X|\ch}$), then the evaluation of the bound can be performed without any knowledge
of the statistical nature  of the interference.

\section{Proofs}\label{sec: proof of rate bound}
\subsection{Proof of Lemma \ref{Lem: rate lower bound}}
Using the Gel'fand and Pinsker result, (\ref{e: costa rate defintion}), an achievable rate is given by:
\begin{IEEEeqnarray}{rCl}\label{e: second rate defintion}
R&=&
h(U| \ch   )-h(U| \ch,Y   )-h(U| \ch   )+h(U| \ch,Z   )
\nonumber \\
&=&
h(U| \ch,Z )-h(U| \ch,Y   )
\TwoOneColumnAlternate{\nonumber \\
&=&}{=} h(U| \ch,Z )-h(U-B Y| \ch,Y   )
\nonumber \\
&\ge& h(U| \ch,Z )-h(U-B Y| \ch   )
\end{IEEEeqnarray}
where $B$ can be any function of the channel, $\ch$. To derive the lower bound, we further limit the discussion to an auxiliary variable that is a linear combination:
\begin{IEEEeqnarray}{rCl}\label{e: linear assignament}
U=X+A Z
\end{IEEEeqnarray}
where $A$ may depend on the channel $\ch$. Thus
we have:
\begin{IEEEeqnarray}{rCl}
R&\ge&  h(X| \ch ,Z)
\TwoOneColumnAlternate{\nonumber \\
&&}-h\left( (1-B\ch)   X +(A-B)Z -B N\Big| \ch   \right).\IEEEeqnarraynumspace
\end{IEEEeqnarray}
Optimizing with respect to the input distribution and the choice of the variables $A$ and $B$, and recalling that the notation $h(X|\ch)$ means $E_\eta[h(X|\ch=\eta)]$, leads to (\ref{e: lemma equation}) and
completes the proof of the lemma.

\subsection{Proof of Theorem \ref{Lem: rate lower bound}}

As we derive a lower bound, we can limit the discussion to any choice of (conditional) input distribution. We find it convenient to choose $X$ to have the distribution of $X_\tG$, i.e., to have a Gaussian
distribution with zero mean and a variance $\sigma_X^2$. Furthermore, $X$ is assumed to be statistically independent of  $Z$ and $\ch$. We write the resulting channel model as:
\begin{IEEEeqnarray}{rCl}\label{d: Gaussain channel model}
Y_\tG = \ch   \cdot X_\tG +Z +N_\tG.
\end{IEEEeqnarray}
  We note also that $\rho_{X_\tG Z}=E [X_\tG (Z-E[Z])]=0$. 

Theorem \ref{Lem: rate lower bound} can be proved directly by optimizing the constants $\alpha$ and $\beta$ in (\ref{e: lemma equation}). A shorter proof can be obtained by first removing the noise correlations.
In order to account for the noise correlation, we rewrite (\ref{d: Gaussain channel model}) as:
\begin{IEEEeqnarray}{rCl}\label{e: revised channel model}
Y_\tG = \tch   \cdot X_\tG + \tilde W Z +\tilde N
\end{IEEEeqnarray}
where:
\begin{IEEEeqnarray}{rCl}\label{e: Ht}
\tch=\ch+\rho_{X_\tG N_\tG}\frac{\sigma_{N_\tG}}{\sigma_X }
\end{IEEEeqnarray}
\begin{IEEEeqnarray}{rCl}\label{e: wt}
\tilde W=1+\rho_{Z N_\tG}\frac{\sigma_{N_\tG}}{\sigma_Z}
\end{IEEEeqnarray}
and
\begin{IEEEeqnarray}{rCl}
\tilde N&=&N_\tG-\rho_{X_\tG N_\tG}\frac{\sigma_{N_\tG}}{\sigma_X} X_\tG
-\rho_{Z N_\tG}\frac{\sigma_{N_\tG}}{\sigma_Z}Z.\IEEEeqnarraynumspace
\end{IEEEeqnarray}
One can easily verify that:
\begin{IEEEeqnarray}{rCl}
E\left[X_\tG \big(\tilde N-E[\tilde N]\big)\right]=E\left[\big(Z-E[z]\big)\big(\tilde N-E[\tilde N]\big)\right]=0. \nonumber
\end{IEEEeqnarray}
Note that the resulting noise variance is:
\begin{IEEEeqnarray}{rCl}\label{e: tilde noise variance}
\sigma_{\tilde N}^2&=\big(1-\rho_{X_\tG N_\tG}^2-\rho_{Z N_\tG}^2\big)\sigma_{N_\tG}^2.
\end{IEEEeqnarray}

Using Lemma \ref{Lem: general lower bound} on the virtual channel of (\ref{e: revised channel model}), and recalling that $X_\tG$ is statistically independent of $Z$ and $\ch$, we have:
\begin{IEEEeqnarray}{rCl}\label{e: rate with entropy for proof}
R&\ge& E\Big[h(X_\tG  )
\TwoOneColumnAlternate{
 \\ \nonumber
&&\hspace{-3mm}}{}-\min_{\alpha,\beta}h\left( (1-\beta\eta)  X_\tG +(\alpha-\beta)\tilde WZ -\beta \tilde N\big| \tch =\eta\right)  \Big].
\end{IEEEeqnarray}
Inspecting (\ref{e: rate with entropy for proof}), the differential entropy in the first line is the entropy of a Gaussian random variable, while the differential entropy in the second line can be upper bounded by the
entropy of a Gaussian random variable with the same variance. Thus, the achievable rate satisfies:
\begin{IEEEeqnarray}{rCl}\label{e: RL lower bound}
R&\ge& \frac{1}{2}\log_2(2\pi e \sigma_X^2)-E\left[\frac{1}{2}\log_2\big(2\pi e Q(\tch)\big)\right]
\end{IEEEeqnarray}
where
\begin{IEEEeqnarray}{rCl}\label{e: define Q}
Q(\eta)&=&\min_{\alpha,\beta}
\mathrm{Var}\left((1-\beta\eta)    X_\tG +(\alpha-\beta)\tilde W Z -\beta \tilde N\right) \nonumber \\
&=&\min_{\alpha,\beta}
(1-\beta\eta)^2    \sigma_X^2 +(\alpha-\beta)^2\tilde W^2 \sigma_Z^2 +\beta^2  \sigma_{\tilde N}^2
\IEEEeqnarraynumspace
\end{IEEEeqnarray}
and the last line of (\ref{e: define Q}) used the fact the $X_\tG$, $Z$ and $\tilde N$ are uncorrelated.
To obtain the tightest bound, we need to minimize $Q(\eta)$ with respect to $\alpha$ and $\beta $. As the parameter $\alpha$ affects only the middle term in the last line of (\ref{e: define Q}), the optimal value of $\alpha$
is obviously $\alpha=\beta $, leading to:
\begin{IEEEeqnarray}{rCl}\label{e: short Q}
Q(\eta)
&=&
\min_{\beta} (1-\beta \eta )^2    \sigma_X^2  +\beta ^2 \sigma_{\tilde N}^2.
\end{IEEEeqnarray}
To find the optimal value of $\beta $, we take the derivative of (\ref{e: short Q}) with respect to $\beta $ and equate to $0$:
\begin{IEEEeqnarray}{rCl}\label{e: deriv Q}
-2\eta (1-\beta \eta )    \sigma_X^2  +2\beta  \sigma_{\tilde N}^2=0
\end{IEEEeqnarray}
resulting with
$\beta =\eta\sigma_X^2 /(\eta ^2 \sigma_X^2  +\sigma_{\tilde N}^2  )
$.

Substituting back in (\ref{e: short Q}) gives the minimal value of:
\begin{IEEEeqnarray}{rCl}\label{e: minimal Q}
Q(\eta)
&=&
\frac{\sigma_{\tilde N}^4\sigma_X^2+\sigma_X^4\sigma_{\tilde N}^2 \eta ^2}{\bigl(\eta ^2 \sigma_X^2  +\sigma_{\tilde N}^2  \bigr)^2}
=\frac{\sigma_{\tilde N}^2\sigma_X^2}{\eta ^2 \sigma_X^2  +\sigma_{\tilde N}^2  }
 .
\end{IEEEeqnarray}
Substituting (\ref{e: minimal Q}) back in (\ref{e: RL lower bound}) results with
\begin{IEEEeqnarray}{rCl}
R&\ge&
E\biggl[
\frac{1}{2}\log_2\biggl(1+\frac{ \tch   ^2 \sigma_X^2}{\sigma_{\tilde N}^2 }\biggr)
\biggr].
\end{IEEEeqnarray}
To complete the proof we need to substitute (\ref{e: Ht}) and (\ref{e: tilde noise variance}),
resulting with (\ref{e: basic theorem}).
\hfill$\blacksquare$

We note that an alternative proof, based on the multi-dimensional lattice approach of Khina and Erez \cite{khina2010robustness}, is presented in \cite{Bergel2013SPAWC} for the special case where the noise
is assumed to be uncorrelated with the input and the interference terms. This alternative proof has the advantage of being a constructive proof, i.e., it presents a coding scheme that can actually achieve the bound,
but it is more restrictive due to the uncorrelation assumption.

\section{conclusions}\label{sec: conclusions}
Lower bounds on the achievable data rates in an additive noise additive interference channel are derived, for the case where the interference is known to the transmitter. It is shown that dirty paper coding is
applicable for this setup even when the noise and the interference are non-Gaussian and/or are mutually dependent on each other and on the input signal. 
Thus, the results indicate that a DPC standard precoder, designed for Gaussian noise and known interference, is robust in more general channel settings.
The relaxed assumptions on the structure of the noise and interference signals facilitates the usage of the bounds for the analysis of actual communications systems. For these, both the known interference as well as the additive unknown noise may be associated with structured signaling, such as transmissions of other users, and hence the Gaussian assumption may not fit.

The bounds are phrased in terms of second order statistics of certain auxiliary variables, and are in the general form of the classical Shannon capacity formula for the interference free AWGN channel. The bounds are tight when the noise is close to Gaussian and can be interpreted
as an extension of the well-known result that the `Gaussian noise is the worst noise' to the more elaborate interference channel model considered in this letter.
\bibliographystyle{IEEEtran}
\bibliography{../../Bergel_all_bib}

\end{document}